\documentclass[aps,floatfix,nofootinbib]{revtex4} 
 
\usepackage{amsmath,amssymb} 
 
\usepackage{mathrsfs} 

\usepackage{epsfig,psfrag} 
\usepackage{array,rotating} 
 
\begin{document} 
\newcommand{\mrs}{\mathscr} 
 
\newcommand{\vecb}[1]{\mbox{\boldmath$#1$}} 
 
\renewcommand{\figurename}{Fig.} 
\renewcommand{\tablename}{Table} 
 
\title{$J$-matrix and Isolated States} 
\author{A. M. Shirokov}\email{shirokov@nucl-th.sinp.msu.ru} 
\address{ Skobeltsyn Institute of Nuclear Physics, Moscow State University, 
Moscow, 119992, Russia}    
\author{ S. A. Zaytsev}\email{zaytsev@mail.khb.ru} 
\address{Physics Department, Khabarovsk State Technical University, 
Tikhookeanskaya 136, Khabarovsk 680035, Russia} 

\begin{abstract}\centerline{\parbox{.92\textwidth}{\hspace{10pt} 
We show that a quantum system with nonlocal interaction can have bound
states of unusual type --- Isolated States (IS). IS is a bound state that is
not in correspondence with the $S$-matrix pole. IS can have a positive
as well as a negative
energy and can be treated as a generalization of the bound states embedded
in continuum on the case of discrete spectrum states. The formation of IS in
the spectrum of a quantum system is studied using a simple rank--2 separable
potential with harmonic oscillator form factors. Some physical applications
are discussed, in particular, we propose a separable $NN$ potential
supporting IS that describes the deuteron binding
energy  and the $s$-wave triplet and singlet scattering phase
shifts. We use this potential to examine the so-called problem of
the three-body bound state collapse discussed in literature. We show
that the variation of  the two-body IS energy causes drastic changes of the
binding energy and of the spectrum of excited states  
of the three-nucleon system. }}
\end{abstract}
\maketitle 
 
\section{Introduction}
Quantum mechanical bound states are known to have the wave functions
decreasing rapidly at large distances $r$. Usually the bound states
are possible at negative energies ($E<0$) only while at positive
energies ($E>0$) the system has only continuum spectrum states with wave
functions oscillating at large distances $r$. Nevertheless von
Neumann and Wigner showed long ago \cite{9} that a quantum system can have a
bound state at positive energy $E>0$. Such states are conventionally
refered to as `{\em continuum bound states}' or as `{\em bound states
embedded in continuum}' (BSEC). 
Von Neumann and Wigner used a local potential in their study 
of BSEC. BSECs are also natural when the interaction is nonlocal (see
\cite{10} and references therein) or in the case of multichannel
scattering (see, e.~g., \cite{Newton,Multichannel,Multichannel2}).

   A  phenomenological nonlocal  nucleon-nucleon ($NN$) potential
   supporting BSEC was suggested by Tabakin \cite{Tabak}. The
potential  predicts the $NN$ data fairly well. However it was found
   that Tabakin potential generates an extremely large binding energy
   for the three-nucleon system \cite{Beam,Aless}. Such a `{\em bound state
   collapse}' was investigated by several groups of workers
\cite{Fiedel,Rupp,Delfino1,Delfino2,Nak-Mae,Sofi} with Tabakin and
   similar nonlocal potentials. All these groups associated the bound state
   collapse with the 
two-body BSEC  and suggested various
   interpretations of the origin of such a puzzling phenomenon.

In some of these studies \cite{ Delfino1,Delfino2,Sofi},  BSEC was
interpreted as an $S$-matrix pole on the real positive energy
axis. This is obviously a mistake since it is well known from
textbooks \cite{Landavshitz} that  the  unitarity condition for 
scattering requires $|S(E)|=1$ for real $E>0$ in the
case of elastic scattering ($|S(E)|\leq 1$ for real $E>0$ in the
case of inelastic scattering), hence the $S$-matrix $S(E)$ cannot have
poles  on the real positive energy axis. BSEC appears to be a very
interesting state: contrary to the conventional discrete spectrum
states it is a bound state which is not associated with any of the
$S$-matrix poles~\cite{Sm}. We introduce {\em Isolated States} (IS)
that are, by 
definition, bound states that do not correspond to the $S$-matrix
poles. Isolated states are a generalization of BSECs: any BSEC is IS, 
however the energy 
$E_I$ of IS
can be also negative, in particular, the ground state of the system
can be an Isolated State. Generally, if the $S$-matrix $S(E)$ is known
than  we obtain the energies of the  discrete
spectrum states by associating the $S$-matrix poles at negative energies with
the discrete spectrum states. The information about the IS energy 
cannot be extracted from the $S$-matrix, such bound states 
are {\em isolated} from the continuum spectrum states. 

In the next Section we present a nonlocal interaction supporting IS. 
Within the $J$-matrix formalism, it is
easy to formulate a realization of this interaction which makes it
possible to find a simple analytical expression for the $S$-matrix. 
This interaction is used to  study the IS formation 
when the Hamiltonian parameters are varied, the IS contribution 
to  the Levinson theorem, etc. 

We address the bound state collapse problem in Section~III. We fit the
parameters of our exactly solvable nonlocal interaction model
supporting IS
to the
$NN$ scattering data. The IS energy 
$E_I$ is arbitrary
because it is not related to the $S$-matrix. We show that 
varying the IS 
energy $E_I$ (note that the $S$-matrix is  unaffected by this variation)
we produce great changes  of the three-body binding
energy: some $E_I$ values bring us to the  $^3$H system with three
bound states with extremely large (few GeV) binding energies; with
larger $E_I$ values we obtain  the  $^3$H nucleus with two bound
states (the binding energy of the ground state is of the order
of few hundreds MeV); the further increase of $E_I$ results in the
further decrease with $E_I$ of the  binding energy of the $^3$H
nucleus which has a single bound state; if the IS energy $E_I$ is
large enough the  $^3$H nucleus becomes unbound. The
experimental value of the $^3$H binding energy can be exactly
reproduced if   some particular positive energy $E_I$ is taken when IS 
appears to be BSEC. Therefore the   problem of the
three-body bound state collapse does not exists as a general problem:
this  problem arose only
due to the use of  very restricted models of interaction supporting BSEC 
for the construction of $NN$ potentials, i.~e. this problem is
inherent for such interaction models only.

We present in Section~IV a short discussion and compare our results
with the results of other authors who studied the three-body bound
state collapse problem.

\section{Simple potential model supporting  Isolated State}

     The radial wave  function  $\Psi^{l}_{E}(r)$  satisfies  the
Schr\"{o}dinger equation
\begin{equation}
(T^l + V^l - E)\Psi^{l}_{E}(r)=0,          \label{(1)}
\end{equation}
where $E$ is the energy, $l$ is the angular momentum, $T^l$ is
the kinetic energy operator and $V^l$ is the potential energy.

Let $\{|i\rangle\}$, $i=0,$ 1, 2, ... be a complete $L^2$ basis. The
Hamiltonian matrix $H_{ij}\equiv\langle i|H|j\rangle$ is generally infinite and
the wave function $\Psi^l_E(r)$ at any energy $E$ can be expressed
generally as an infinite expansion in basis functions,
\begin{equation}\label{Psi}
\Psi^l_E(r) =\sum_{i=0}^\infty \alpha_i(E)\,|i\rangle.
\end{equation}

However, at some particular energy $E_I$ the infinite Hamiltonian
matrix $H_{ij}$ can have a finite eigenvector, and the
wave function $\Psi_I(r)$  at this energy is expressed as a finite
expansion in basis functions, 
\begin{equation}\label{PsiI}
\Psi_I(r) =\sum_{i=0}^M \alpha_i(E_I)\,|i\rangle.
\end{equation}
The wave function $\Psi_I(r)$ rapidly decreases with distance $r$ since
it is a superposition of a finite number of $L^2$ functions. Therefore
at energy $E_I$ we have a bound state of a particular type hereafter
refered to as an {\em Isolated State} (IS). 

Clearly, we have the IS solutions of the type (\ref{PsiI}) in the case
when  the Hamiltonian matrix $H_{ij}$ is block-diagonal, 
\begin{equation}
\label{block-d}
H_{ij}=H^{(1)}_{ij}\oplus H^{(2)}_{ij},
\end{equation} 
where the $(M+1)\times (M+1)$ submatrix $H^{(1)}_{ij}$ is defined in a
finite-dimensional subspace spanned by
the basis functions $|i\rangle$ with $i=0,\,1,\,...\,,\,M$. 
The infinite-dimensional
submatrix $H^{(2)}_{ij}$ is defined in the orthogonal supplement to
this subspace. Any eigenvector of the submatrix $H_{ij}$ gives rise to
the wave function of the type (\ref{PsiI}), i.~e. each of  the
submatrix $H_{ij}$ eigenvectors is associated with IS.

The scattering state  wave functions
 with oscillating asymptotics at energies $E>0$, can be
expressed only as a superposition of an infinite number of $L^2$
functions, 
\begin{equation}
\label{PsiS}
\Psi^l_E(r) =\sum_{i=M+1}^\infty \alpha^{(2)}_i(E)\,|i\rangle,
\end{equation}
where $\left\{\alpha^{(2)}_i(E)\right\}$ are eigenvectors of the  infinite
submatrix $H^{(2)}_{ij}$. The $S$-matrix and scattering phase shifts are
defined through the asymptotics of the functions (\ref{PsiS}), hence they are
governed by the structure of the submatrix $H^{(2)}_{ij}$ only. The
energy $E_I$ of IS and its other features  are 
dictated by the structure of the other submatrix  $H^{(1)}_{ij}$ that
is generally independent from  $H^{(2)}_{ij}$. Therefore we cannot
expect that the $S$-matrix has a pole at the IS energy 
$E_I$. We can define the Isolated State as the bound state that
is not associated with any of the $S$-matrix poles. The IS energy 
$E_I$ can be positive, and in this case it appears to be the
so-called {\em bound state embedded in continuum} (BSEC). 
Any BSEC is IS. 
The IS energy 
$E_I$ can be also negative, in particular, the ground state of the system
can be isolated. Thus IS 
can be treated as a
generalization of BSEC on the case of arbitrary (negative or positive) energy.

The $J$-matrix formalism \cite{2} makes it possible to study IS
properties 
and to formulate a simple exactly-solvable model of a
system  possessing IS. 
In this contribution we use the oscillator basis; the
exactly-solvable model of IS can be also easily formulated by means of
the $J$-matrix formalism with the Laguerre basis.

The idea of the block-diagonal structure of the Hamiltonian matrix
(\ref{block-d}) can be easily  realized if the
interaction between the particles is  
described by a separable nonlocal potential of the rank $N+1$,
\begin{equation}
{V}^l = \sum_{n,n'=0}^{N} V^l_{nn'}\, 
|\varphi_{nl}(r)\rangle\langle\varphi_{n'l}(r')| , 
          \label{(2)}
\end{equation}
with  the harmonic oscillator form factors
\begin{equation}
\varphi_{nl}(r) = (-1)^{n} \left[ \frac{2n!}{r_{0}\Gamma (n+l+\frac{3}{2})}
\right] ^{\frac{1}{2}} \left( \frac{r}{r_{0}} \right) ^{l+1} \exp \left(
- \frac{r^2}{2r^{2}_{0}} \right) L_{n}^{l+\frac{1}{2}} \left(
\frac{r^2}{r_{0}^{2}} \right).
\label{(3)}
\end{equation}
Here $r_{0} =(\hbar /m\omega )$ is the oscillator radius and
$L^{\alpha}_{n} (x)$ is the Laguerre polynomial. 

In the $J$-matrix method,
the wave function has a form of 
series in terms of $L^{2}$
functions (\ref{(3)}), 
\begin{equation}
\Psi ^{l}_{E} (r) = \sum_{n=0}^{\infty} X_{n} (E)\: \varphi_{nl}(r).
                 \label{(4)}
\end{equation}
The coefficients $X_{n} (E)$ 
for $n\geq N$ are given by the formula
\begin{equation}
X_{n} (E) = S_{nl}(p)\, \cos \delta _{l} + C_{nl}(p)\, \sin \delta_{l}
\; ,                                               \label{(5)}
\end{equation}
where $p=\sqrt{2E/\hbar\omega}$  is the momentum, $S_{nl}(p)$ and
$C_{nl}(p)$ are the eigenvectors of the infinite tridiagonal matrix of the
kinetic  energy
$T^l_{nn'}$.   The following
analytical expressions \cite{2} can be used to calculate 
$S_{nl}(p)$  and  $C_{nl}(p)$: 
\begin{align}
&S_{nl}(p) = \left[ \frac{2\Gamma (n+l+\frac{3}{2})}{\Gamma (n+1) } \right]
^{\frac{1}{2} } \frac{p^{l+1} }{\Gamma (l+\frac{3}{2}) } \exp
(-\frac{p^2}{2}) \: _{1}F_{1}(-n,\, l+\frac{3}{2};\, p^{2})
\: ,                                              \label{(6a)} \\[1ex]
&C_{nl}(p) = \left[ \frac{2\Gamma (n+1)}{\Gamma (n+l+\frac{3}{2}) }
\right] ^{\frac{1}{2} } \frac{(-1)^{l} }{p^{l} \Gamma (-l+\frac{1}{2}) }
\exp (-\frac{p^2}{2})\;
 _{1}F_{1}(-n-l-\frac{1}{2},\, -l+\frac{1}{2};\, p^{2})\:
.                                                \label{(6b)}
\end{align}
The phase shift $\delta _l$  in the
partial wave with the angular momentum $l$ can be calculated as
\begin{equation}
\tan \delta _{l} = - \frac{S_{Nl}(p) - \wp _{NN}(E)\,S_{N+1,l}(p)}
{C_{Nl}(p) - \wp _{NN}(E)\,C_{N+1,l}(p)}, 
         \label{(9)}
\end{equation}
where
\begin{equation}
\wp _{nn'}(E) = - \sum_{\mu } \frac{U^{\mu }_{n}\: U^{\mu }_{n'} }
{\varepsilon _{\mu} - E } T^{l}_{n',n'+1} ,               \label{(8)}
\end{equation}
$U^{\mu}_{n}\; (n = 0,1,...,N)$ are the eigenvectors and 
$\varepsilon_{\mu}$ are the corresponding eigenvalues of the truncated
Hamiltonian  matrix
$\tilde{H}^{N}_{nn'} =T^{l}_{nn'} + V^{l}_{nn'} \;
(n,n'=0,1,...,N)$.
The coefficients $X_{n}(E)$ for $n\leq N$ can be found by the formula
\begin{equation}
X_{n}(E) = \wp _{nN}(E)\,X_{N+1}(E).
                     \label{(7)}
\end{equation}

\begin{figure}
\unitlength=0.80mm
\special{em:linewidth 0.4pt}
\linethickness{0.4pt}
\begin{center}
\begin{picture}(72.34,74.33)
\put(22.67,29.33){\framebox(29.67,30.00)[cc]{$H^{(2)}$}}
\put(7.67,59.33){\framebox(15.00,15.00)[cc]{$H^{(1)}$}}
\put(52.34,29.33){\line(1,-1){20.00}}
\put(48.34,29.33){\line(1,-1){20.33}}
\put(52.34,33.33){\line(1,-1){20.00}}
\put(14.01,46.66){\makebox(0,0)[cc]{{\large 0}}}
\put(37.34,67.33){\makebox(0,0)[cc]{{\large 0}}}
\put(37.34,17.33){\makebox(0,0)[cc]{{\large 0}}}
\put(66.67,42.66){\makebox(0,0)[cc]{{\large 0}}}
\end{picture}
\end{center}\vspace{-5ex}
\caption{The structure of the Hamiltonian matrix.}
\label{Ham-str}
\end{figure}
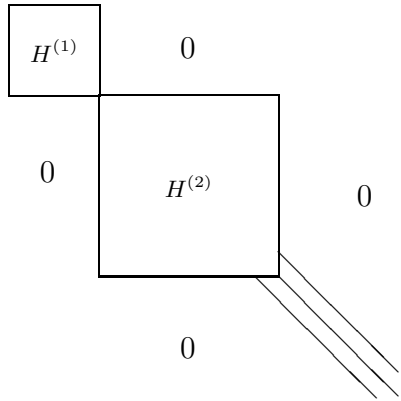

We are considering the case when the Hamiltonian matrix is
block-diagonal. We note that the kinetic energy matrix in the
oscillator basis is tridiagonal. Hence with  the interaction
(\ref{(2)}) we can obtain  the Hamiltonian matrix in the
oscillator basis of the type (\ref{block-d}) that has the structure
shown in Fig.~\ref{Ham-str}.
%
Solid lines schematically show the
infinite tridiagonal kinetic energy
tail of the Hamiltonian matrix, the rest non-zero matrix elements
are displayed by two boxes representing submatrices $H^{(1)}$
and $H^{(2)}$ (we include  the  infinite  tridiagonal  kinetic
energy tail in the submatrix $H^{(2)}$).

Obviously, the eigenvectors of the submatrix $H^{(1)}$ are the
eigenvectors of the  entire Hamiltonian matrix $H$, 
too. The corresponding wave functions are decreasing with~$r$
similarly to the bound state wave functions  since
they are superpositions (\ref{PsiI}) of a finite number of the
oscillator functions $|i\rangle=\varphi_{nl}(r)$ given by  Eq.~(\ref{(3)}).
However these  wave functions 
have an unusual asymptotics
$\sim\!\exp [-{r^{2}}/(2r^{2}_{0})]$ instead of the standard one
$\sim\!\exp(-\alpha r)$. All eigenstates of the submatrix
$H^{(1)}$ are ISes. It is obvious that generally the
submatrix $H^{(1)}$ may have positive or/and negative eigenvalues
$\varepsilon ^{(1)}_{\mu}$. If $\varepsilon ^{(1)}_{\mu} > 0$, the
corresponding state appears to be BSEC. If $\varepsilon ^{(1)}_{\mu} < 0$,
the corresponding state appears to be a bound state of a specific type.

The continuum spectrum states with the oscillating asymptotically  wave
functions as well as the conventional bound states with the
$e^{-\alpha r}$--type  wave function
asymptotics, can be expressed only as infinite series (\ref{PsiS})
of the oscillator functions $|i\rangle=\varphi_{nl}(r)$ given by
Eq.~(\ref{(3)}).  
They are generated by the submatrix $H^{(2)}$. The scattering
and  conventional bound state wave functions~(\ref{PsiS}) are
obviously orthogonal to the IS  wave functions~(\ref{PsiI}). Thus 
the scattering and usual bound state
wave functions have node(s)  at  a
small distance ($\sim$$r_0$) in the presence of IS(es)  similarly
to the wave  functions generated by the potentials with forbidden
states~\cite{6}. 

The asymptotic behavior of the scattering state wave functions is characterized
completely by the $S$-matrix \cite{Newton}. Hence the structure of the
$S$-matrix is governed by
the infinite-dimensional submatrix $H^{(2)}$ that is independent from the
submatrix $H^{(1)}$. The IS energies $\varepsilon ^{(1)}_{\mu}$, on the other
hand, are controled by the submatrix $H^{(1)}$ and are independent from the
submatrix $H^{(2)}$. Varying matrix elements $H^{(1)}_{nn'}$ of the
submatrix $H^{(1)}$ one causes variation of the IS energies
$\varepsilon ^{(1)}_{\mu}$ without affecting the $S$-matrix. Thus,
the energy of IS $\varepsilon ^{(1)}_{\mu}$ is not in correspondence with
the location of the $S$-matrix poles. Using symmetry properties
of the $S$-matrix as
a function of the complex momentum $p$ \cite{Newton}, it is easy to
show \cite{Sm} 
that the energy of BSEC is not in correspondence with any of the $S$-matrix
poles. An interesting new point, so far as we know never discussed in
literature, is the appearance of the discrete spectrum
states, i.~e., of ISes with negative energy $\varepsilon ^{(1)}_{\mu}<0$, that
are divorced from the $S$-matrix poles. So, IS being the state with
the asymptotically decreasing wave function and with the energy at which
the $S$-matrix does 
not have a pole, can been treated as a generalization of BSEC on the case of
the discrete spectrum states.

    We  examine in more detail the formation of IS in the spectrum of
a quantum system with nonlocal interaction using as an example a simple
analytically solvable model. The simplest realization of the situation
depicted in Fig.~\ref{Ham-str}, corresponds to the case when the
submatrix $H^{(1)}$ is a $1\times 1$ matrix and the separable potential
(\ref{(2)})  is
of the rank 2, i.~e. $N=1$. In this  case,   IS arises due to
the cancellation of the potential  energy  matrix  elements
$V_{01} = V_{10}$  and the kinetic energy  matrix  elements
$T_{01} = T_{10} = -V_{01}$ that results in $H_{01} = H_{10} = 0$. The
IS energy $\varepsilon _{0}$  is  equal  to  the  diagonal  matrix
element $H_{00}$, $\varepsilon _{0} = H_{00}$. It should be stressed
that $H_{00}$ can  take
an arbitrary value in our model.  Using
Eqs. (\ref{(6a)}--\ref{(8)}) we obtain the following expression for the
phase shift $\delta _l$:
\begin{equation}
\tan \delta _{l} =
 - \frac{S_{0l}(p) \left\{ \left[
V_{11}(\varepsilon _{0} - E) - \beta  \right] (T^l_{00} - E)
+ \left(T^{l}_{01}\right)^2 (\varepsilon _{0} - E ) \right\} }
{C_{0l}(p) \left\{ \left[
V_{11}(\varepsilon _{0} - E) - \beta  \right] (T^l_{00} - E)
+ \left(T^{l}_{01}\right)^2 (\varepsilon _{0} - E ) \right\} -
\frac{p[V_{11}(\varepsilon _{0} -E) - \beta ]}{\pi S_{0l}(p) } } ,
   \label{(tan)}
\end{equation}
where $\beta \equiv H^{2}_{01}$.

\begin{figure}
\psfrag{D}{\begin{sideways}$\delta_0$, degrees\end{sideways}}
\centerline{\psfig{figure=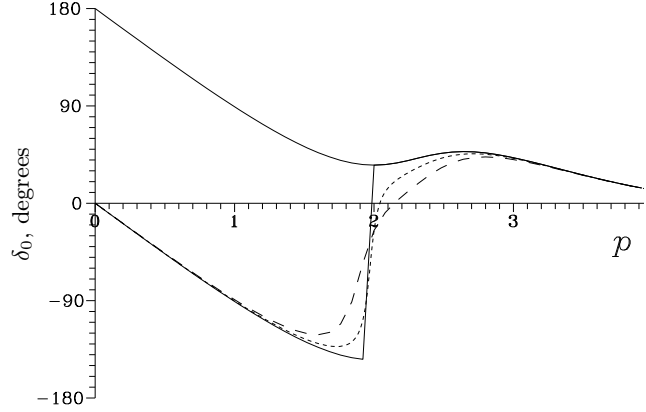,width=.7\textwidth}}
\caption{ The evolution of the phase shift $\delta_{0}(p)$   as
$\beta = H_{01}^2 = H_{10}^2$ tends to zero. Dashed, dotted, and solid
curves are the phase shifts obtained with  $\beta=\beta_1$, $\beta_2$
and $\beta_3$, respectively;
$\beta_{1} > \beta_{2} > \beta_{3} = 0$.}
\label{evol}
\end{figure}

    Suppose  $\varepsilon _{0} >0$  and $V_{11} > 0$.
The  evolution  of  the  $s$ wave  phase  shift $\delta_{0}(p)$   as
$\beta = H_{01}^2 = H_{10}^2$ tends to zero is shown in  Fig.~\ref{evol}.
If $\beta \neq 0$
there is a resonance at the energy $E \approx \varepsilon_0$
of  the  width  $\Gamma$ that  decreases as $\beta$ is reduced.
When $\beta =0$,
BSEC arises as  the
resonance of the zero width producing the jump of the height $\pi$ of  the
phase shift $\delta_0$   at the energy $E=\varepsilon_0$.
This spurious jump should be eliminated
that results in the $\delta_{0}(0)$  increase by an extra $\pi$ if we
suppose, as usual, that $\delta_0(\infty)=0$. Thus when
applying the Levinson theorem \cite{Newton} to the system pertaining
IS, one should count 
IS as a usual discrete spectrum bound state. Such behavior of the
phase  shift  is  typical  for systems pertaining BSEC that have been
studied in various models \cite{Newton,Sm,7}. Thus our model represents an
alternative simple analytical approach in the study of BSEC. 

     As for the S-matrix, it is given by the following expression:
\begin{equation}
S_{l} =
 - \frac{C^{(-)}_{0l}(p) \left\{ \left[
V^l_{11}(\varepsilon _{0} - E) - \beta  \right] (T^l_{00} - E)
+ \left(T^{l}_{01}\right)^2 (\varepsilon _{0} - E ) \right\}
-\frac {p[V^l_{11}(\varepsilon _{0} -E) - \beta ]}{\pi S_{0l}\vphantom{^l}(p)}}
 {C^{(+)}_{0l}(p) \left\{ \left[
V^l_{11}(\varepsilon _{0} - E) - \beta  \right] (T^l_{00} - E)
+ \left(T^{l}_{01}\right)^2 (\varepsilon _{0} - E ) \right\} -
\frac{p[V^l_{11}(\varepsilon _{0} -E) - \beta ]}{\pi S_{0l}\vphantom{^l}(p)}},
                            \label{(10)}
\end{equation}
where $C^{(\pm )}_{nl}(p)=C_{nl}(p) \pm iS_{nl}(p)$. The single
$S$-matrix pole on the unphysical sheet tends
to  the  real  energy   $E=\varepsilon_0$  as $\beta$ tends to  zero.
However  in  the
limit $\beta =0$, the factor $(\varepsilon_{0} -E)$
in the numerator of Eq.~(\ref{(10)}) cancels the same
factor in the denominator and  the  singularity  at the energy
 $E=\varepsilon_{0}$  disappears:
\begin{equation}
S_{l} =
 - \frac{C^{(-)}_{0l}(p)\left[
V^l_{11}(T^l_{00} - E) + \left(T^{l}_{01}\right)^2\right]
-\frac {pV^l_{11}}{\pi S_{0l}\vphantom{^l}(p) }}
 {C^{(+)}_{0l}(p) \left[V^l_{11}(T^l_{00} - E)
+ \left(T^{l}_{01}\right)^2 \right] -
\frac{pV^l_{11}}{\pi S_{0l}\vphantom{^l}(p) }} ,
                            \label{(10-2)}
\end{equation}
This illustrates the mechanism of the $S$-matrix pole loss in the limit
$\beta\to 0$ when the resonance transforms into BSEC.
The nontrivial result is that if IS is a bound state  
($\varepsilon_{0} <0$),
it  does not  generate   the $S$-matrix pole, too.

     The above results can be easily generalized
as the following statement.

{\em
Let all eigenvalues of the truncated Hamiltonian
matrix $H^N$  be non-degenerate. Then  Isolated  States  occur  in  the
spectrum of the system with the rank--$(N+1)$ separable  interaction
(\ref{(2)})
if and only if the truncated matrix $\tilde{H}^N$   and  its  principal  minor
$\tilde{H}^{N-1}$  of the rank $(N-1)$ have common eigenvalues.
The  number $\nu$ of
the common eigenvalues is equal  to  the  number  of  the Isolated
States. These eigenvalues  and  the corresponding  eigenfunctions  are
just the energies and the wave functions of the  Isolated  States. }


     The equivalent formulation 
is:

{\em   The system with the nonlocal interaction (\ref{(2)}) has the Isolated
State at the energy  $\varepsilon_{\mu}$  if
and only  if  $\varepsilon_{\mu}$ is the eigenvalue  of the
truncated Hamiltonian matrix
$\tilde{H}^N$ and the corresponding
eigenvector $U^{\mu}$ has the last component $U^{\mu}_{N} = 0$.}

\section{Phenomenological $NN$ potential with Isolated State and
the three-nucleon system}


The simplest rank-2 separable potential (\ref{(2)}) supporting IS
discussed in the previous Section, was
used to fit the $NN$ singlet ${^1s_0}$ and triplet  ${^3s_1}$
scattering phase shifts.  The oscillator function parameter
$\hbar\omega=500$~MeV. To ensure the existence of IS, we should set 
the off-diagonal
potential energy matrix elements $V_{01}=V_{10}=-T_{10}$. The phase
shifts are independent of the  matrix
element $V_{00}$ that governs the IS energy. The only parameter
responsible for the phase shifts is the matrix element $V_{11}$.  

\begin{figure}
\vspace{-1.2cm}
\psfrag{D}{\begin{sideways}$^{1\!}\delta_0$, degrees\end{sideways}}
\psfrag{E}{$E_{\rm lab}$, MeV}
\centerline{\epsfig{file=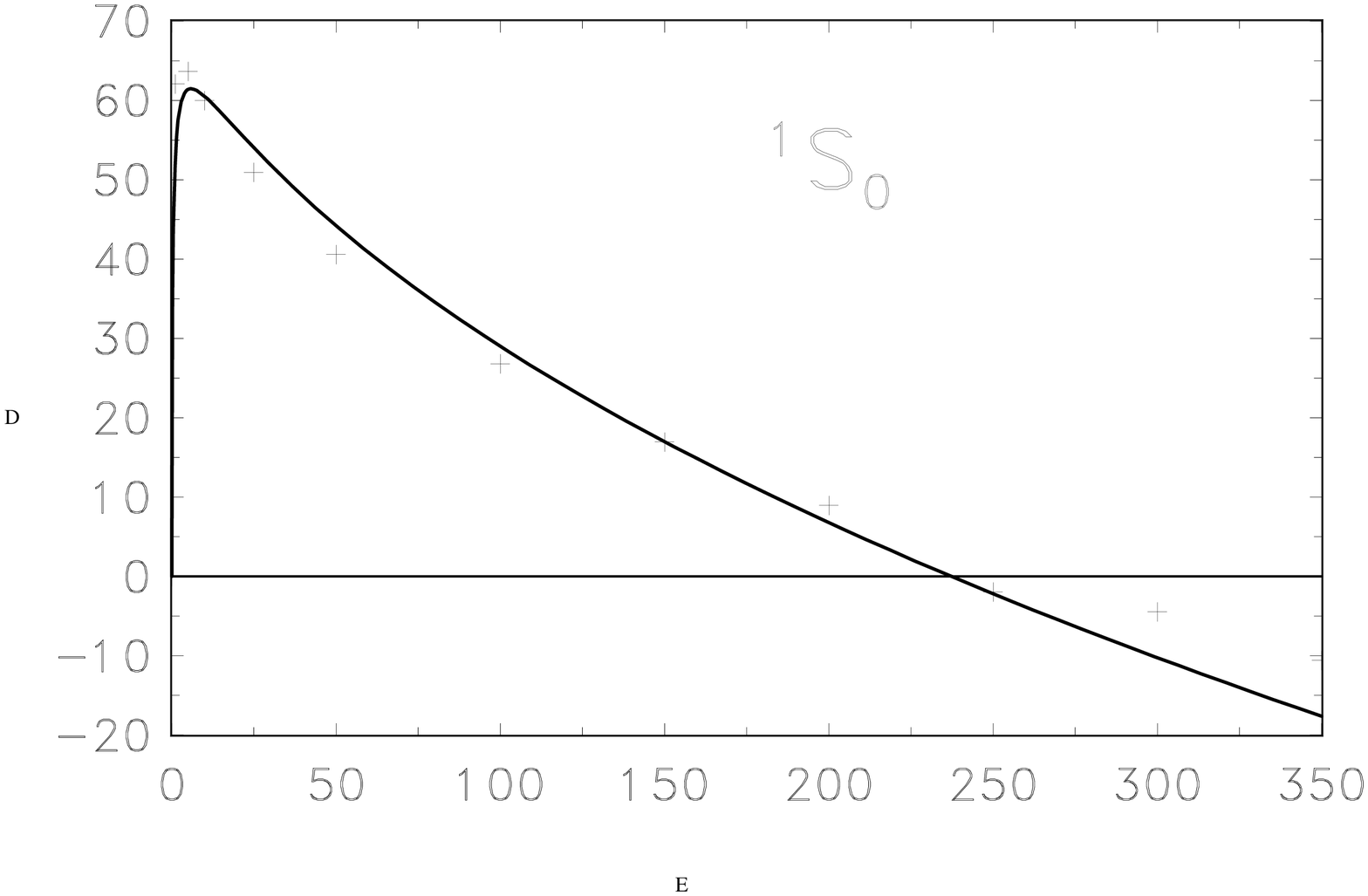,width=.8\textwidth}}
\vspace{-1.cm}
\caption{Singlet $s$ wave phase shifts. Solid line --- phenomenological
potential with IS; $+$~--- experimental data.}
\label{S-S}

\vspace{-.3cm}
\psfrag{D}{\!\!\!\begin{sideways}\!\!\!\!$^{3\!}\delta_1$, 
                                        degrees\end{sideways}}
\psfrag{E}{$E_{\rm lab}$, MeV}
\centerline{\epsfig{file=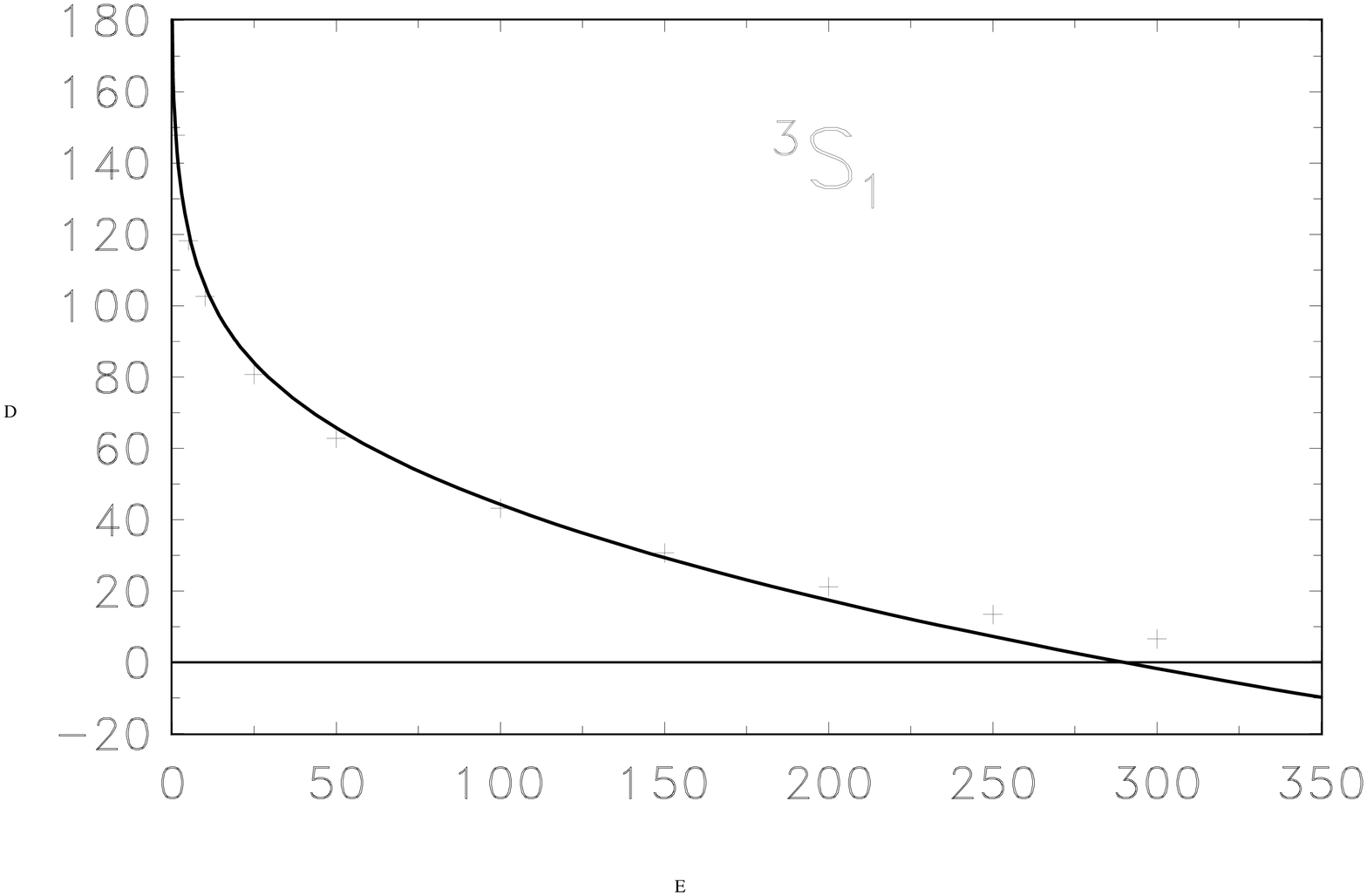,width=.8\textwidth}}
\vspace{-1.cm}
\caption{Triplet $s$ wave phase shifts. Solid line --- phenomenological
potential with IS; $+$~--- experimental data.}
\label{S-T}
\end{figure}

The singlet phase shifts obtained with $V^s_{11}=-0.7315\,\hbar\omega$
and the triplet  phase shifts obtained with
$V^t_{11}=-0.81512\,\hbar\omega$ are shown in Figs.~\ref{S-S} and
\ref{S-T} respectively.  These values of $V^s_{11}$ and  $V^t_{11}$
are seen to reproduce with a reasonable accuracy the scattering data.

The deuteron ground state energy should be obtained by the calculation
of the $S$-matrix pole as is discussed in Refs.~\cite{Halo-b,Istp-c}. 
Since in our case the $S$-matrix is given by the
expression~(\ref{(10-2)}), the $S$-matrix poles can be calculated by
solving the equation
\begin{equation}
 {C^{(+)}_{0l}(p) \left[V^l_{11}(T^l_{00} - E)
+ \left(T^{l}_{01}\right)^2 \right] -
\frac{pV^l_{11}}{\pi S_{0l}\vphantom{^l}(p) }} =0 .
                            \label{(10-Eq)}
\end{equation}
The bound state (deuteron) should be searched for in the triplet ${^{3\!}s_1}$
wave,  i.~e. $l=0$ and  $V^t_{11}$ should be used as $V^l_{11}$ in
Eq.~(\ref{(10-Eq)}). We are searching for  negative $E_d$
value and $p_d=i\sqrt{2|E_d|/\hbar\omega}$ fitting
Eq.~(\ref{(10-Eq)}). The deuteron wave function can be calculated
using the $J$-matrix formalism (see also the discussion in
Refs.~\cite{Halo-b,Istp-c} and references therein). Our
very simple potential with the only fitting parameter $V^t_{11}$
provides a good description of the deuteron energy
$E_d=-2.22496$~MeV and rms radius $\sqrt{\langle r^2\rangle}=1.87$~fm
(the respective experimental values are $E_d^{\rm exp}=-2.224575$~MeV
and  $\sqrt{\langle r^2_{\rm exp}\rangle}=1.9676$~fm). 

\begin{figure}
\vspace{-1.cm}
\psfrag{Et}{\begin{sideways}\!\!\!\!$E_t$, Mev\end{sideways}}
\psfrag{Ei}{$E_I$, MeV}
\centerline{\epsfig{file=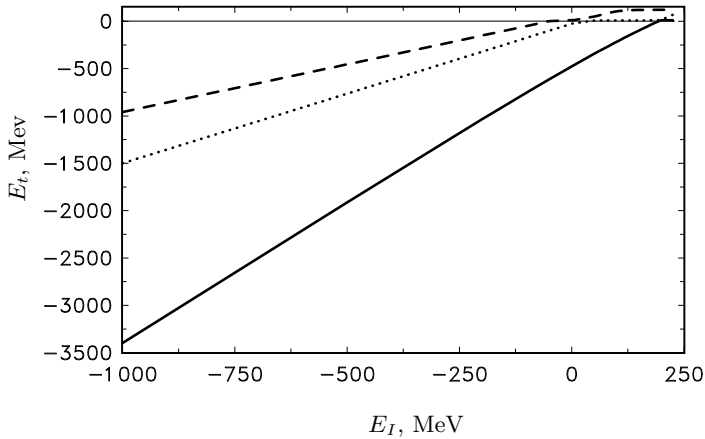,width=.8\textwidth}}
\vspace{-.4cm}
\caption{The triton ground and excited  state energies vs the IS
energy $E_I$.}
\label{IS-T}
\end{figure}

The obtained singlet and triplet $s$ wave $NN$ potentials are used in
the calculation of the triton bound states. As in the other studies of the
three-body bound state collapse 
\cite{Beam,Aless,Fiedel,Rupp,Delfino1,Delfino2,Nak-Mae,Sofi},
we do not allow for the interaction in other partial waves. We perform
a conventional variational calculation of the $S=T=J=\frac12$
three-nucleon states
with the three-body oscillator
basis allowing for all components with the total number of oscillator
quanta $N=2n+l\leq 32$. The energies of the ground state and of the
lowest excited states are shown in Fig.~\ref{IS-T} for different $E_I$
values (we are varying both the triplet $V^t_{00}$ and  the singlet
$V^s_{00}$ potential energy matrix elements so that
$V^s_{00}=V^t_{00}$; in other words the IS energy $E_I^t$ in the
triplet state is equal to   the IS energy $E_I^s$ in the singlet
state: $E_I=E_I^s=E_I^t$). It is seen that
the variation of  the IS energy $E_I$ (that does not affect the phase
shifts and the deuteron properties) results in the drastic changes of
the triton binding energy and of the spectrum of excited
$S=T=J=\frac12$ states.  When the IS energy $E_I$ is small enough, the
three-nucleon system collapses (the binding energy becomes extremely
large); two excited states are bound in 
addition to the ground state.
The triton ground state energy increases with the IS energy $E_I$; the
same is true for the energies of the excited states. At some $E_I$ value 
the second excited state becomes unbound; at some larger 
$E_I$  value the first excited state becomes unbound, too; however the
triton binding energy is still too large. Nevertheless the further
increase of $E_I$ results in the increase of the triton ground
state energy and at some $E_I$ value the three-nucleon system becomes
unbound. 

We see that we really have the three-body bound state collapse for
some values of the IS energy $E_I$. However the collapse disappears
for larger $E_I$ values. Therefore
our general conclusion is that the three-body bound state collapse
problem does not exist as a general problem. In the previous studies
of the three-body bound state collapse, very restricted potential models
were used that did not allow to change the BSEC energy $E_{\rm BSEC}$.
The $E_{\rm BSEC}$ value used in these studies caused the extreme
overbinding of the trinucleon. This is clearly the problem of the
particular potentials used and not the general problem of the
three-body bound state collapse inherent for all two-body potentials
supporting BSEC. In our model any trinucleon binding energy can be
obtained by phase equivalent variation of the IS energy $E_I$. In
particular, we can fit the $E_I$ value to the triton binding
energy. Setting $E_I=E_I^s=E_I^t=189.525$~MeV, we obtain
the triton binding energy of $E_t=8.47307$~MeV
in our $32\hbar\omega$ variational calculations.
The accuracy of the variational approach is improved if the three-body
$S$-matrix pole is calculated as is discussed in Ref.~\cite{Halo-b}:
in this case  the triton binding energy of $E_t=8.4748$~MeV is obtained.

\section{Discussion}

We suppose that the most important feature of BSEC is that this bound
state is not in correspondence with the $S$-matrix pole. We introduce
IS as a generalization of BSEC on the case of arbitrary (positive or
negative) energy. We propose a simple model of non-local interaction
elucidating the formation of IS. To  investigate in detail the
formation and the features of IS,
it is natural to employ the
$J$-matrix formalism  that makes it possible to obtain analytical
expressions for the $S$-matrix and other observables, i.~e. to
formulate the exactly solvable model of IS.

The general results given  above  can  be straightforwardly extended
on  the coupled channel  case,  
on the case of the separable finite-rank nonlocal potentials
with the Laguerre form factors , etc.

It is interesting that IS naturally appeares in the standard nuclear
shell model. 
It is well-known (the Dubovoy-Flores theorem) \cite{DF,11}
that the lowest eigenvalue $\varepsilon_{0}^{N}(\omega )$  
obtained in the shell model calculation in the $N\hbar\omega$ model
space, 
coincides with the lowest eigenvalue
$\varepsilon_{0}^{N-2}(\omega )$    obtained in the shell model
calculation in the $(N-2)\hbar\omega$ model  space 
if the parameter $\hbar \omega$
of the oscillator basis  minimizes
the eigenvalue $\varepsilon_{0}^{N-2}(\omega )$.
According to the statement formulated in Section~III,   
the ground state obtained in the conventional variational nuclear
shell-model calculations appears to be IS.


Using our exactly solvable model of IS, we obtain a simple nonlocal
$NN$ potential describing  the singlet and
triplet $s$ wave phase shifts and deuteron properties. With this $NN$
interaction we study the
triton properties and examine the three-body bound state collapse
problem. We can vary the IS energy $E_I$ in our model without
violating the description of the phase shift and deuteron properties,
i.~e. without altering
the on-shell properties of the interaction. However the off-shell
properties of the interaction are modified when $E_I$ is varied and
this is clearly seen in the strong $E_I$-dependence of the triton
ground state energy $E_t$. The three-body system collapses at some
$E_I$ values  and does not at other  $E_I$ values. We can even fit the
position of IS to the triton binding energy. Therefore we
conclude that the discussion of the three-body bound state collapse in
literature \cite{Fiedel,Rupp,Delfino1,Delfino2,Nak-Mae,Sofi} arose only 
due to the use of unsuccessful $NN$ potential supporting BSEC; other
two-body interactions supporting BSEC provide a correct description of
the three-nucleon binding.

Various interpretations of the nature of the three-body bound state
collapse have been suggested in literature
\cite{Fiedel,Rupp,Delfino1,Delfino2,Nak-Mae,Sofi}. We cannot accept
most of them. 

For examples,  Nakaichi-Maeda \cite{Nak-Mae} and Pantis et al
\cite{Sofi} supposed that the three-body bound state collapse arises
due to the nodel behavior of the deuteron wave function in the case of
potentials supporting BSEC. Delfino et al \cite{Delfino2} supposed
that collapse originates from the structure of
the BSEC wave function which is identical with that of the Pauli
forbidden state.
In our model, we shifted the IS energy
changing neither the deuteron wave function nor the IS wave
function. We show that the  three-body bound state collapse can
be eliminated by the increase of the IS energy only, therefore 
we cannot agree that
the collapse originates from the structure of the deuteron or IS wave
function. 

Delfino et al \cite{Delfino1} supposed that the three-body bound state
collapse is a manifestation of the Thomas effect (the increase of the
triton binding energy when the range of the $NN$ potential tends to
zero). The range of the general nonlocal interaction cannot be established
unambiguosly. Delfino et al suggested to use the average
kinetic energy $\langle T\rangle$ of the two-body bound state as an
indicator  of the potential range and demonstrated that BSEC is accompanied
by an increase of $\langle T\rangle$. We cannot agree with this
conclusion. In our model the bound state wave functions are unaltered
when the IS energy is varied, hence the variation of IS energy does
not cause changes of   $\langle T\rangle$ of any of the bound
state. Therefore with the same values of  $\langle T\rangle$ of
two-body bound states, we obtain either the collapsed triton at some $E_I$
values or normally-bound (or just unbound) triton at other $E_I$
values.

Rupp et al \cite{Rupp} note that the
triton binding energy increases when the BSEC energy grows. We obtain
an opposite result in our investigation: the triton binding energy
decreases with the BSEC energy. We note here that in our studies we
shift BSEC phase equivalently while Rupp et al base their
conlusion on the results obtained with a number of potentials
providing different phase shifts. In other words, they use potentials
with different on-shell properties that can mask any effect of the
BSEC energy variation.

It is interesting that it is easy to change the BSEC (or any other
bound state) energy phase equivalently. The BSEC wave function
$\Psi_{\rm BSEC}$ fits the Schr\"odinger equation
\begin{equation}
\label{BSEC}
H\Psi_{\rm BSEC}=E_{\rm BSEC}\Psi_{\rm BSEC}.
\end{equation}
Let us define a new Hamiltonian
\begin{equation}
\label{H'}
H'=H+\lambda|\Psi_{\rm BSEC}\rangle\langle\Psi_{\rm BSEC}|.
\end{equation}
The wave functions
$\Psi_E^l(r)$ of 
all the rest (bound or continuum spectrum) eigenstates
of the Hamiltonian $H$, fit the Schr\"odinger equation 
\begin{equation}
\label{rest}
H\Psi_{E}^l(r)=E\Psi_E^l(r).
\end{equation}
Clearly $\Psi_E^l(r)$ will be also the eigenfunctions of the
Hamiltonian $H'$ since $\Psi_{\rm BSEC}$ is orthogonal to any
$\Psi_E^l(r)$. At the same time, the BSEC energy $E_{\rm BSEC}$ will
be increased by $\lambda$. We are sure that if Rupp et al employed
this method of varying the BSEC energy, they would obtain the natural
result that the triton binding energy decreases with  $E_{\rm BSEC}$.

The Hamiltonian (\ref{H'})  can be defined  with the projector 
$|\Psi_{E_b}^l\rangle\langle\Psi_{E_b}^l|$ on any other bound state
$\Psi_{E_b}^l(r)$ at the energy $E_b$. In this case we obtain another
simple model of IS. Really, none of the continuum spectrum wave
functions $\Psi_E^l(r)$ is altered by the projector  
$|\Psi_{E_b}^l\rangle\langle\Psi_{E_b}^l|$. Hence the Hamiltonians $H$
and $H'$ provide the same $S$-matrix. Thus the projector  
$\lambda|\Psi_{E_b}^l\rangle\langle\Psi_{E_b}^l|$ increases the energy
$E_b$ of the bound state by $\lambda$ but does not shift the
$S$-matrix pole at the negative energy $E_b$. As a result the
conventional bound state at the energy $E_b$ is transformed into
IS. We note here that the Hamiltonians with the projection
operators  $\lambda|\Psi_{E_b}^l\rangle\langle\Psi_{E_b}^l|$ are
widely used, e.~g., to project out Pauli forbidden states (see, e.~g.,
\cite{Kukulin}).

We do not suppose that it is needed to criticize other interpretations
of the origin of the three-body bound state collapse suggested in
various papers. Our general conclusion is that the collapse is
inherent for very restricted models of potentials supporting BSEC
only and does not appear if other two-body interactions with BSEC are
used. Therefore for the proper interpretation of the three-body bound
state collapse origin, one  should carefully study this restriction
and reveal what is wrong with it. Up to the best of our knowledge, it
was never done.

\bigskip 

This work was supported in part  by the State Program ``Russian
Universities'' and by the Russian Foundation of Basic research, Grant 
No~02-02-17316.


\end{document}